# Necessity of Two More Space Lattices

M. A. Wahab, Deptt. of Physics, Jamia Millia Islamia,

New Delhi-110025, India.

Abstract

Existing fourteen Bravais/space lattices are found to be inadequate for proper crystallographic representation and some conceptual understandings. The complicacies, confusions and ambiguities encountered in the representation of trigonal/rhomohedral and hexagonal lattices have been removed by suggesting the need of two more space lattices in the form of Hexagonal Close Packing (HCP) and Rhombohedral Close Packing (RCP) with appropriate justifications. Henceforth, these two lattices will be called as Wahab lattices and the sixteen lattices together as Bravais–Wahab lattices or simply B–W lattices (or space lattices as before). This finding will have immediate as well as far reaching implications.

## 1. Introduction

The complications, confusions and ambiguities in the representation of trigonal/rhombohedral and hexagonal lattices are found to be widely reported in the literature.(1) Suggestion of the fifteen lattices is seldom found in the literature without proper justification.(1)(2) A thorough literature survey on this aspect clearly shows the disagreement over the actual number of space lattices that are possible in three dimensions. As a result of this confusion, some important crystallographic concepts (for example, 73 symmorphic space groups cannot be derived on the basis of 14 space lattices) could not be understood properly.

In order to remove the above mentioned confusions and ambiguities, a thorough investigation related to the macroscopic symmetries have been made. The analysis of the investigation led to conclude the necessity of two more lattices, namely the HCP and RCP lattices (henceforth named as Wahab lattices) as the independent members of the space lattices for a complete solution of the persisting problem. A brief description of the aspects that are causing complicacies and confusions is reviewed. Suitable justifications have been provided for the selection of HCP and RCP as independent lattices. Their inclusion in the list of space lattices help understand the missing crystallographic data.

## 2. Complications and Confusions Related to Representations

### (i) Hexagonal and Trigonal/Rhombohedral Crystal Systems

A survey of presently available literature indicates the existence of complications and confusions in the representation of hexagonal and trigonal/rhombohedral lattices.(1) For example, the quartz crystal is supposed to have a hexagonal lattice, whereas it does not have hexad symmetry but have triad symmetry. Such crystals are assigned to the trigonal system rather than to the hexagonal system. Trigonal system is also treated as a special case of the hexagonal system since both have the same relationships between the unit cell axes.(3) In some cases, in place of trigonal system a rhombohedral system is used so that the crystal systems remain seven. The difference between the trigonal and the rhombohedral is also not exactly clear. Further, it is often found that the trigonal system includes crystals of both hexagonal and rhomohedral lattices.

### (ii) Hexagonal Close Packing (HCP) and Cubic Close Packing (CCP)

The discussions and illustrations related to the stacking of close packed planes (layers) in HCP and CCP structures are prevalent in the literature.(4) The stacking of close packed planes in HCP is considered along [001] direction as shown in Fig.1a. On the other hand, the stacking of close packed planes in CCP/FCC is considered along [111] direction as shown in Fig.1b. The abrupt change in the direction (a tilt by an angle of 45° w.r.t. horizontal or vertical) of packing in the two close packed systems seem to be quite arbitrary.

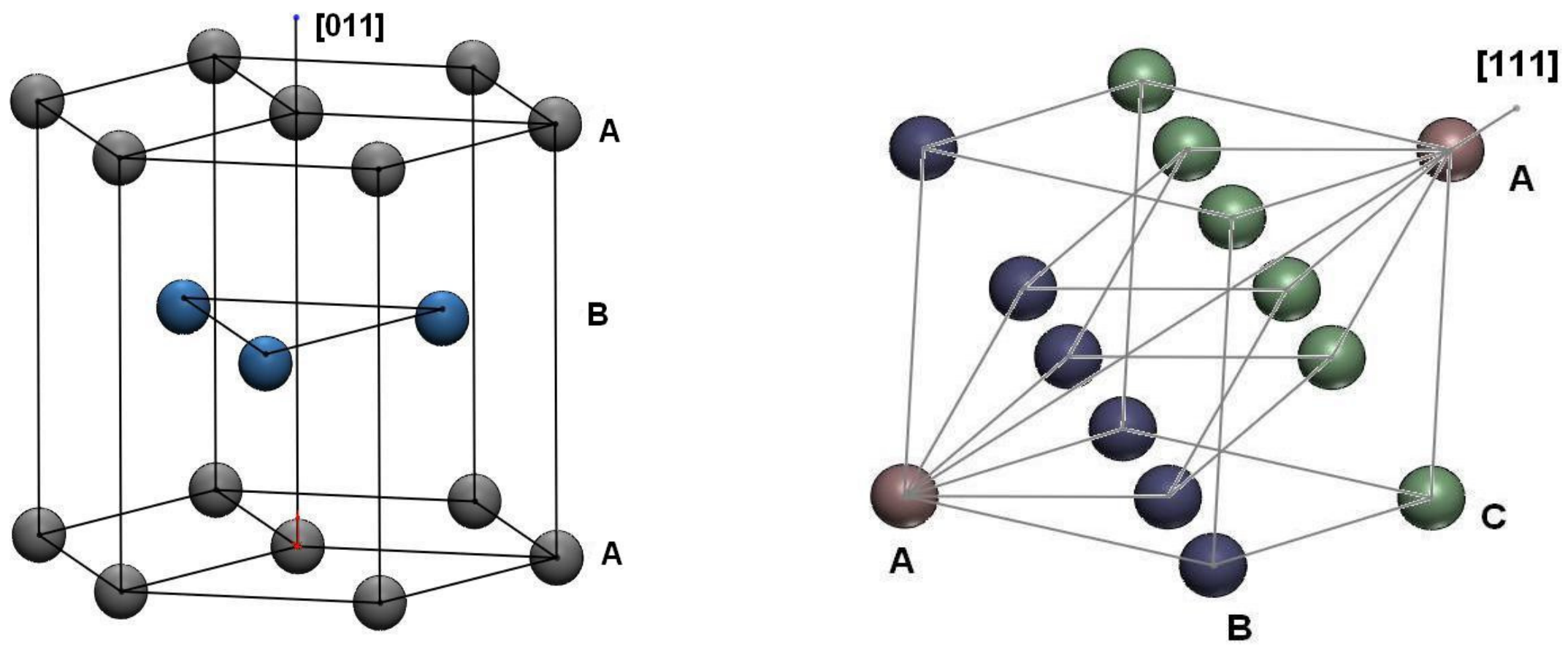

(a) (b)

Fig. 1 (a) HCP unit cell (b) FCC/CCP unit cell

**(iii) Primitive and Non-Primitive Lattices**

There are two categories of lattice that we know. They are primitive lattice and non-primitive (or centered) lattice. Different types of centering have different names such as base centering, face centering, body centering and rhomohederal centering.(1) However, the representation of HCP unit is an exception. This structural unit is represented in terms of a primitive hexagonal lattice with one basis containing two atoms (one at the origin 000 and the other at $\frac{2}{3}\ \frac{1}{3}\ \frac{1}{2}$) instead of representing it as a centered lattice in consonance with monclinic, orthorhombic, tetragonal and cubic lattices.

**3. Discrepancies in the Crystallographic Data**

The available crystallographic data regarding the number of lattices, point groups and symmorphic space groups (obtained by multiplying the number of lattices with the number of point groups in a given crystal system) related to macroscopic symmetries of one, two and three dimensions have been investigated and analyzed. Important information that are causing complicacies and confusions are summarized in Table 1. A close examination reveals the following discrepancies:

(i) The number of point groups is observed to be double the number of lattice types in one and two dimensions, while in three dimensions this is not so (see columns 2A and 3 of Table 1).

(ii) Correspond to 14 space lattices, the number of possible symmorphic space groups that have been derived is 61 as given in column 4A of Table 2, while the correct (maximum possible) number of symmorphic space groups in three dimensions is 73.(5,6)

**4. Understanding of the Problems**

**(i) Formation of SH, HCP and CCP Structures**

To understand the formation of Simple Hexagonal (SH) hexagonal close packed (HCP) and cubic close packed (CCP) structures, let us consider the arrangement of equal spheres of radius R on a triangular lattice as shown in Fig. 2a. The symmetry of this layer about the central sphere (surrounded by six neighboring spheres) is 6mm. When another identical layer is placed on the top of this layer such that all the upper spheres just touch the tips of the lower spheres, a simple hexagonal unit cell is obtained as shown in Fig. 2b. The symmetry of this unit cell becomes 6/mmm. The packing efficiency of a simple hexagonal structure is found to be 60%, widely reported in literature.[7]

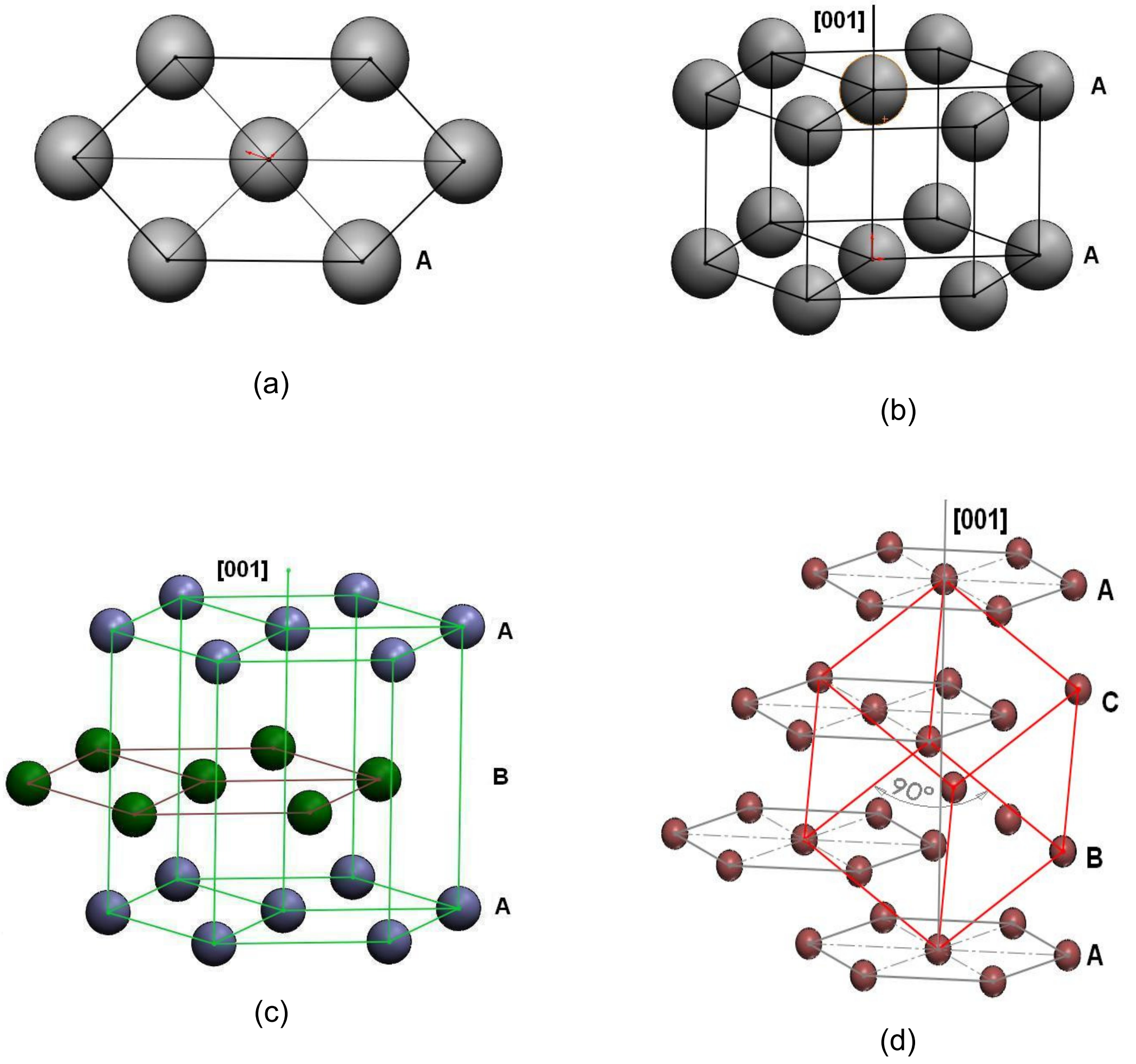

Fig. 2 Formation of (b) SH, (c) HCP and (d) RCP (CCP) from (a) 2-D HCP layer

In order to make the HCP structure, the second layer of spheres is displaced w.r.t the reference layer such that the spheres just fit into say B voids. Similarly, the third layer of spheres is placed such that the spheres lie on the top of the reference layer. The resulting HCP unit cell containing three close packed layers in the sequence ABA…, one on the top of the other along [001] direction is shown in Fig. 2c. The reported symmetry of this unit is $6_3$/mmc and the packing efficiency is estimated to be 74%, also widely reported in the literature.[7]

In order to make the CCP structure, we proceed upto the second layer in the same manner as in the HCP structure and the third layer is also similarly displaced w.r.t the second layer such that the spheres just fit into say C voids. Finally, the fourth layer is placed just on the top of the reference layer. The resulting CCP unit cell containing four close packed layers in the sequence ABCA…., one on the top of the other along [001] direction (and not along [111] direction as per the existing convention) as shown in Fig. 2d. The symmetry of this unit is the same as that of an FCC unit cell (Fig. 8) and the packing efficiency is estimated to be 74% in the following section.

### (ii) Formation of Trigon/Rhombohedron within Some Known Unit Cells

In this section, let us discuss the formation of trigon/rhombohedron inside unit cells of known lattice parameters, viz., inside the simple hexagonal unit cells, cubic close packed unit cell stacked along [001] direction (equivalently FCC unit cell stacked along [111] direction) and body centered cubic (BCC) unit cells, respectively.

For the first case, let us consider a layer of identical spheres as shown in Fig. 2a. Now let us stack four such layers, one on the top of the other as in simple hexagonal case to obtain a four layered structure consisting of three simple hexagonal units as shown in Fig. 3. Next, consider the central sphere of the bottom layer as the reference sphere and construct a trigonal/rhombohehedral unit cell by joining the reference atom with three alternate atoms of the first layer, then to the other set of three atoms of the third layer and finally to an atom in the fourth layer lying just above the reference atom as shown in the figure. When the spheres touch each other, the angle of the trigon/rhombohedron is found to be maximum and is equal to 75.52°. This angle is said to be the critical angle $\alpha_c$. However, the angle of the trigon/rhombohedron will decrease as the separation between the hexagonal layers is increased (i.e. as the c dimension of the hexagonal unit cell is increased). The volume of the rhombohedron is given by

$$V_{T/R} = a_R^3\sqrt{1-3\cos^2\alpha+2\cos^3\alpha}$$

From figure, we have $\alpha$=75.52°, $a_R = 2\sqrt{2}\,R$, the number of atoms in the trigonal/rhombohedral unit cell is 3 (one from the corners and two from the vertical axis) showing that it is a non-primitive (or centered) unit cell. Substituting these values, the volume becomes

$$V_{T/R}(SH)=16\sqrt{2}\,R^3\sqrt{1-3\cos^2 75.52+2\cos^3 75.52}$$

$$= 0.9185\times16\sqrt{2}\,R^3$$

The packing efficiency is

$$\text{Efficiency} = \frac{3\times\text{Vol.of one sphere}}{\text{Vol.of unit cell}}$$

$$= \frac{3\times\frac{4}{3}\pi R^3}{0.9185\times16\sqrt{2}R^3} = \frac{\pi}{0.9185\times4\sqrt{2}} = 60\%$$

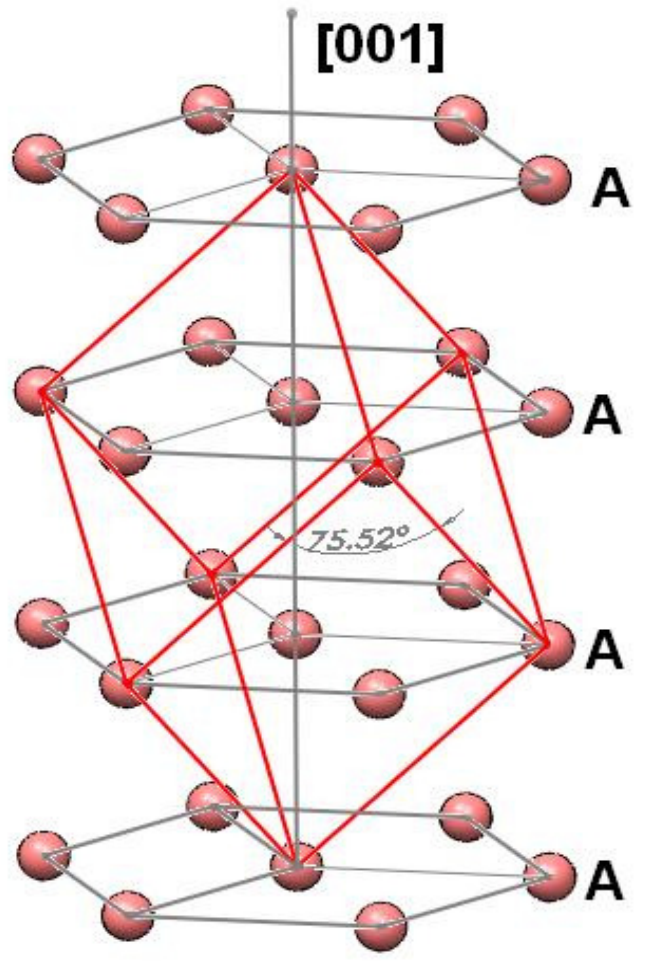

Fig. 3 Formation of a rhombohedron inside SH unit cells

This packing efficiency is the same as that of its parent structure (simple hexagonal system) and hence is better called as simple rhombohedral (SR) or simply trigonal.

Now, in order to form the trigon/rhombohedron in CCP structure in which the layers are stacked along [001] direction, we start with Fig. 2d. Next, construct a rhombohedron within the CCP unit cell by joining the reference atom of the bottom layer with three face centered atoms of the cube lying on the B layer, then to the three face centered atoms lying on the C layer and finally to an atom on the fourth layer (A) lying just on top of the reference atom as shown in Fig. 4. When all the spheres touch each other, the angle of the rhombohedron is found to be minimum and is equal to 60°. This angle is unique in the sense that the structure ceases to remain close packed as soon as the angle is increased. The volume of the trigon/rhombohedron is given by

$$V_{T/R} = a_R^3\sqrt{1\text{-}3\cos^2\alpha+2\cos^3\alpha}$$

From figure, we have α=60°, $a_R$=2R and the number of atom in the unit cell is one only from the corners showing that it is a primitive unit cell. Substituting these values, the volume of the cell becomes

$$V_{T/R}\,(RCP) = 8R^3\sqrt{1\text{-}3\cos^2 60+2\cos^3 60}$$

$$= 4\sqrt{2}R^3$$

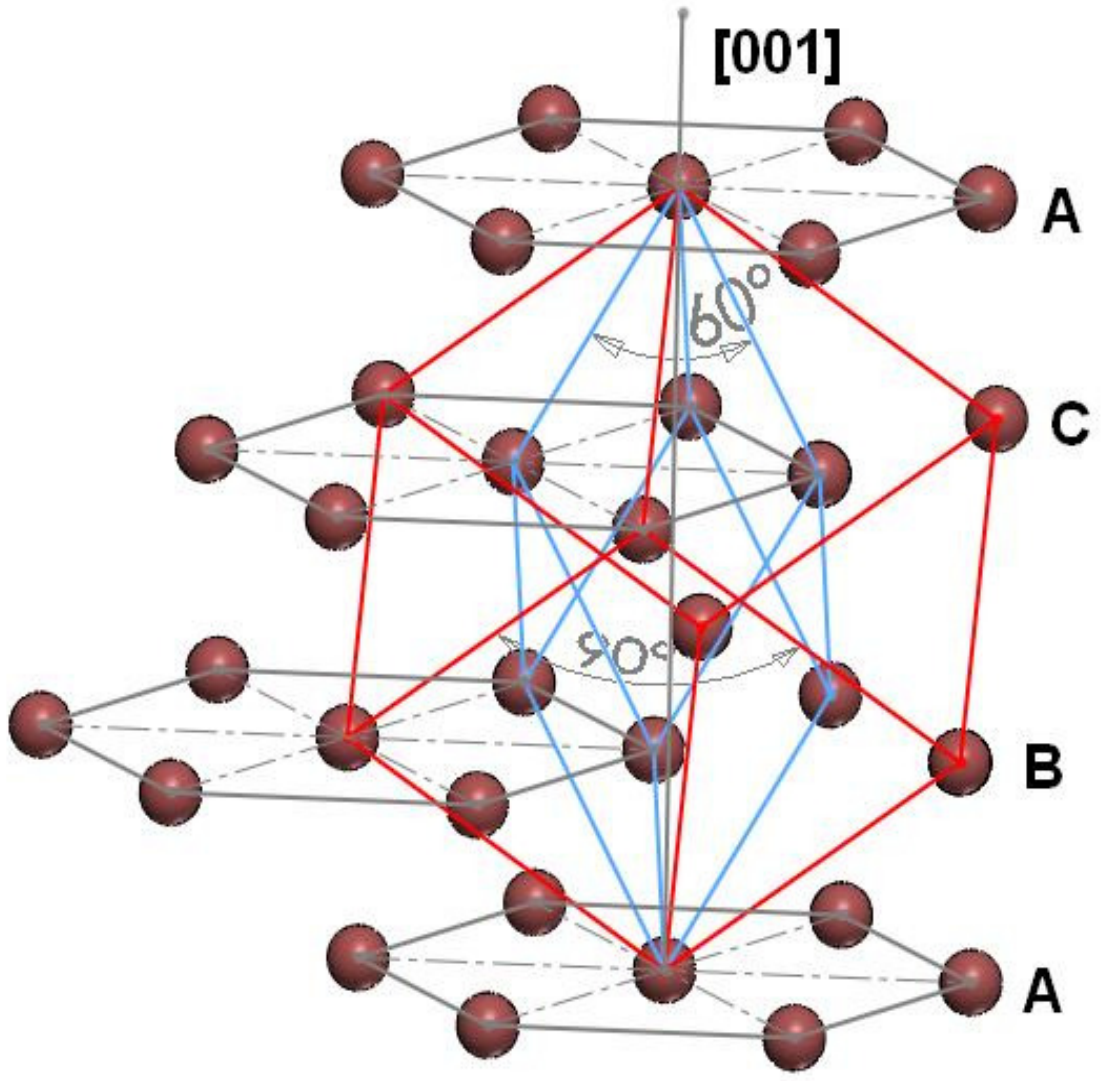

Fig. 4 Formation of a rhombohedron inside CCP unit cell

The packing efficiency is

$$\text{Efficiency} = \frac{\frac{4\pi}{3}R^3}{4\sqrt{2}R^3} = \frac{\pi}{3\sqrt{2}} = 74\%$$

This packing efficiency is the same as that of HCP or its parent structure CCP and for this reason this particular rhombohedron is better called the rhombohedral close packing (RCP). It is important to note that the above consideration is identical to the formation of a rhombohedron inside an FCC unit cell along [111] direction. Further, from the frequent observation of rhombohedral structures in close packed polytypic compounds [(8)] and the uniqueness of its angle (α=60°), one can easily conclude that the rhombohedral structure is the monopoly of close packing only.

Let us consider the case of formation of a trigon/rhombohedron inside BCC unit cells. For the purpose, consider three BCC unit cells side by side in one plane and join their body centered points with one corner as shown in Fig. 5. They represent the side of the trigon/rhombohedron. Now join these points with other appropriate points to obtain the required trigon/rhombohedron. When the BCC condition $\sqrt{3}a = 4R$ is satisfied, the angle of the trigon/rhombohedron is found to be 109.47°. The volume of the trigon/rhombohedron is given by

$$V_{T/R} = a_R^3\sqrt{1\text{-}3\cos^2\alpha+2\cos^3\alpha}$$

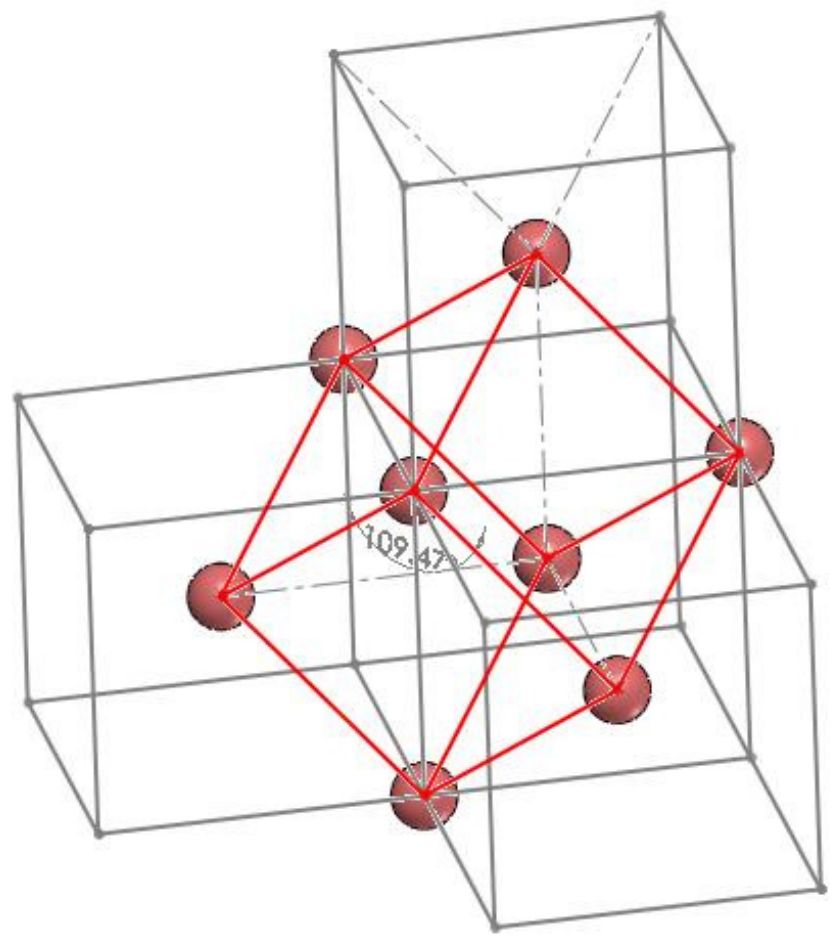

Fig. 5 Formation of a rhombohedron inside BCC unit cells

From figure, we have α=109.47°, $\sqrt{3}a = 4R$ and $a_R = \frac{\sqrt{3}}{2}a = \frac{\sqrt{3}}{2} \times \frac{4R}{\sqrt{3}} = 2R$, so that $a_R^3 = 8R^3$.

Substituting these values, the volume of the cell becomes

$$V_{T/R}\ (\text{BCC}) = 8R^3\sqrt{1\text{-}3\cos^2 109.47+2\cos^3 109.47}$$

$$= 0.77{\times}8R^3$$

The packing efficiency is

$$\text{Efficiency} = \frac{\frac{4\pi}{3}R^3}{0.77{\times}8R^3} \quad = \frac{\pi}{0.77{\times}6} \quad = 68\%$$

This efficiency is the same as that of its parent structure BCC.

The information obtained from the above calculations is provided in Table 3.

**(iii) Projection of SR, HCP and RCP on Basal Plane**

Let us consider the basal plane projection of simple rhombohedron SR i.e. trigon/rhombohedron drawn inside four layers of simple hexagonal nets (Fig. 3), HCP structural unit (Fig. 2c) and the RCP (Fig. 4) i.e. the trigon/rhombohedron drawn inside four cubic close packed hexagonal nets as shown in Fig. 6. The plane projection of the atoms associated with the simple rhombohedron (trigon) exactly falls on the hexagonal positions (three alternate positions each from second and third layer) and gives rise to a primitive hexagonal unit cell (Fig. 6a). On the other hand, the similar plane projections of both HCP and RCP give rise to centered hexagonal unit cells as shown in Figs. 6b and 6c. The position of one centered atom in HCP case is $\frac{2}{3}\ \frac{1}{3}\ \frac{1}{2}$ and two centered atoms in RCP case are $\frac{2}{3}\ \frac{1}{3}\ \frac{1}{2}$ and $\frac{1}{3}\ \frac{2}{3}\ \frac{1}{2}$, respectively.

Here it is important to mention that similar to the existing representation of HCP structural unit, the RCP unit cell could also have been represented as a primitive hexagonal lattice with one basis containing three atoms, one is at the origin and the other two are at $\frac{2}{3}\ \frac{1}{3}\ \frac{1}{2}$ and $\frac{1}{3}\,\frac{2}{3}\,\frac{1}{2}$, respectively. However, this is not possible, because this unit cell is the result of rhombohedral (or cubic) close packing (this is one extreme form of close packing) and can be regarded as an exclusive and independent RCP lattice. However, due to its similar shape it should remain a member of trigonal crystal system. On the other hand, HCP is also another well known extreme form of close packing. Therefore, on a similar ground the HCP should be regarded as an exclusive and independent lattice and should not be clubbed with SH, which is a non close packed structural unit cell. However like RCP, the HCP (because of its similar shape) should continue to be the member of hexagonal crystal system.

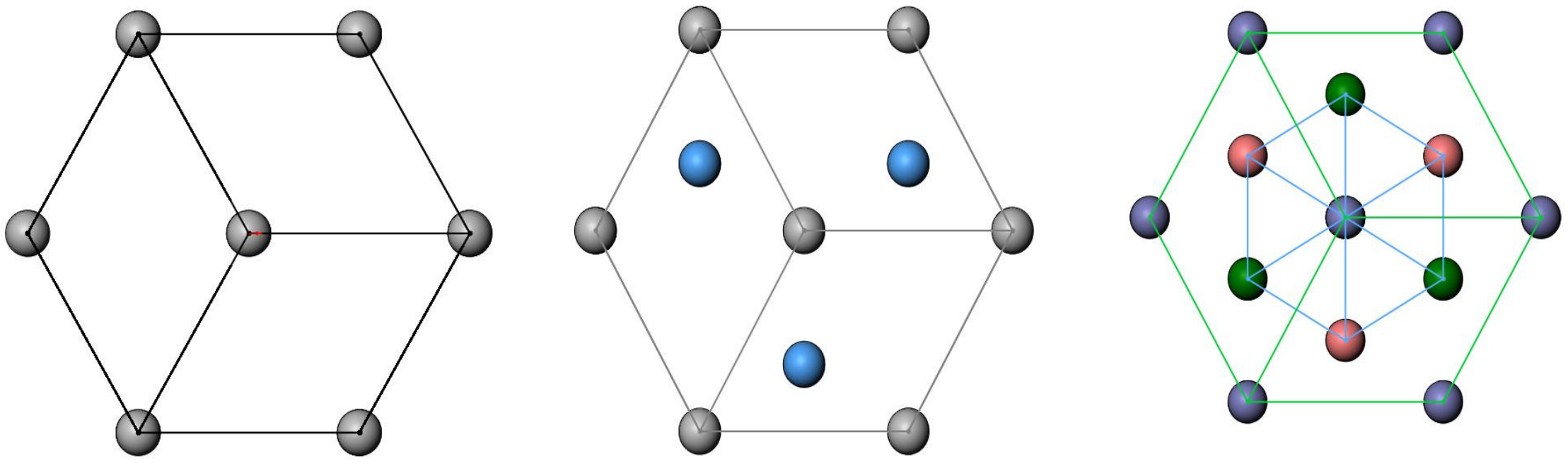

(a) (b) (c)

Fig. 6 Basal plane projection of rhombohedron/trigon in (a) SH, (b) HCP and (c) RCP

### (iv) (111) Plane Projection of BCC and FCC

Let us consider the (111) plane projection of the trigon/rhombohedron drawn inside the BCC and FCC unit cells as shown in Fig. 7. The projection of BCC unit cell is a simple hexagon while that of FCC is equivalent to the projection of RCP on its basal plane. Lines show the cube edges and face diagonals.

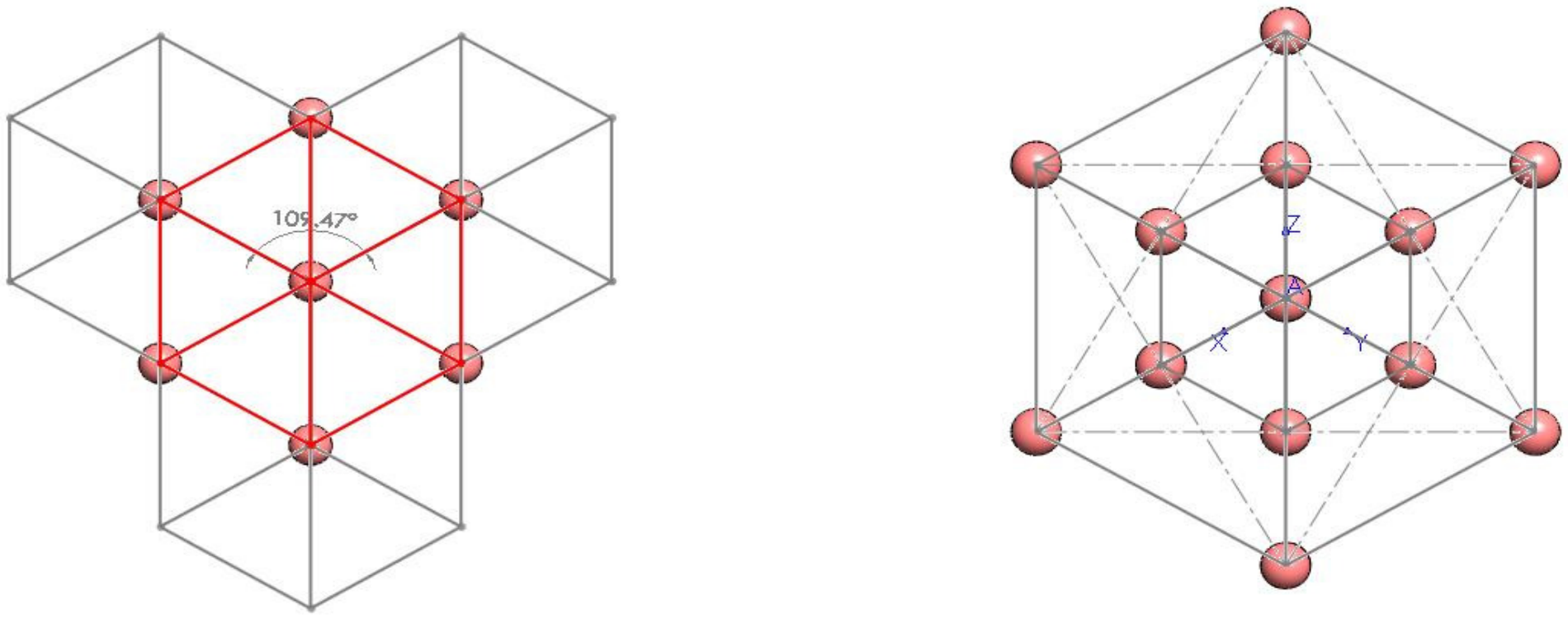

(a) (b)

Fig. 7 (111) plane projection of trigon/rhombohedron drawn inside (a) BCC and (b) FCC unit cells, x, y, z are cubic axes

## 5. Distinction between a Trigon and a Rhombohedron

The literature survey provides us only an ambiguous picture of the trigon and the rhombohedron. However, from the above calculations related to angles and packing efficiencies of trigon/rhombohedron drawn in different cases and their corresponding plane projections provide us two distinct cells. One category showing the plane projections of trigon/rhombohedron constructed within SH and BCC unit cells results in a simple (primitive) hexagonal unit cell as shown in Figs. 6a and 7a with three alternate positions each from second and third layer. Consequently, they can be called as a trigonal or a simple rhombohedral (SR) lattice where α varies in the range 60° < α < 120° excluding α = 90°. On the other hand, the projection of the other (only RCP in Fig. 4) showing a doubly centered hexagonal unit cell (Fig. 6c) which can be called as a rhombohedral close packed (RCP) lattice with α = 60°. This is a unique lattice and needs to be treated separately. The distinct consideration of SR (or trigonal) and RCP lattices makes the total number of space lattices as 16, which clearly removes all the ambiguities and confusions related to their representation and the relationships between the number of lattices with the point groups and symmorphic space groups, respectively (Tables 1 and 2).

### 6. Distinction between an FCC and a CCP

Like the case of trigon and rhombohedron, there exist some ambiguities in the representation of FCC and CCP form of structural units. Sometimes FCC and CCP are used as synonyms. However, the term FCC appears to be used more frequently than the term CCP. We know that an FCC lattice belongs to cubic crystal system with a = b = c and α = β=γ=90°. For an FCC unit cell, $\sqrt{2}a = 4R$ (where R is the radius of the sphere) condition has to be satisfied. The direction of close packing in an FCC unit cell is along [111] direction (Fig. 1b) tilted through an angle of 45° w.r.t. horizontal or vertical. On the other hand, a CCP structure is obtained by placing the close packed layers in the sequence ABCA…, one on the top of the other along [001] direction which is the same as that of HCP as shown in Fig. 2d.

On comparison, we observe that FCC and CCP are indistinguishable except for the direction of packing in mono atomic solids dealing with the packing of identical atoms as in elements. However, for diatomic and above cases (triatomic etc.), FCC and CCP are clearly distinguishable. For example, ZnS, SiC, etc. are close packed structures and belong to CCP while NaCl is an FCC structure but not close packed. A similar argument is applicable for polyatomic cases. Thus a CCP structure will necessarily be a close packed structure while FCC structure may or may not be close packed structure depending on whether it is mono atomic or poly atomic. This suggests that FCC and CCP forms of atomic packing need to be considered separately as far as the representation of non close packed and close packed structures of poly atomic systems are concerned.

Since, [001] is the only direction of close packing in all close packed structures (as in polytypes), the wrongly assigned [111] direction to CCP or RCP structure of polyatomic cases needs to be corrected. As a result of this, the letter F (representing face centered cubic) appearing before the space groups of cubic close packed structures needs to be replaced by either RCP or CCP, as the case may be.

Let us consider one example of close packed polymorphic/polytypic transformation, say from 2H to 3C in terms of layer displacement mechanism as explained in the literature to substantiate the argument given above in relation to the direction of close packing [8] , i.e.

2H : AB AB AB / A

↓ ↓

3C : AB CA BC / A

Here, 3 units of 2H have been considered for transformation. There is no change in the first 2H unit, the second 2H unit changes from AB to CA through layer displacement taking place in a cyclic order like A → B → C → A, while the third 2H unit changes from AB to BC through layer displacement taking place in an anti cyclic order like A → C → B → A. During the above transformation, all layer displacements are taking place in the same horizontal plane and no change of any direction has taken place. The same is valid for all polytypic transformations. They all indicate that the direction of close packing remains the same throughout in all polytypes including HCP and CCP. Table 4 provides a list of some low period hexagonal and corresponding rhombohedral polytypes. A rhombohedron can be oriented into two different ways known as obverse and reverse settings. However, from geometrical point of view, both settings represent the same structure. Since RCP is the smallest possible rhombohedral structure which has three layers in its unit cell, the corresponding hexagonal unit will have only one layer (one-third of the rhombohedral) in it, meaning thereby that in close packing the existence of 1H polytype cannot be ruled out.

## 7. Discussion.

Under different sections, we have tried to pin point the root cause of complications, confusions and ambiguities in the representation of trigonal/ rhombohedral and hexagonal lattices on one hand and the FCC and CCP lattices on the other hand and provided suitable explanation in each case. With the help of various geometrical constructions and projections, we have been able to clearly identify:

1) the difference between a trigon and a rhombohedron
2) the difference between an FCC and a CCP
3) the exclusive and independent identity of HCP and RCP space lattices

The calculation of angle α for trigon/rhombohedron constructed within the CCP, SH and BCC unit cells (whose lattice parameters a,b,c and α,β,γ are more or less defined) have been made while satisfying their structural conditions. The angle α is found to be increasing in the order CCP to BCC and is provided in Table 3. The value α = 60° is found to be a characteristic feature of close packing. For the other two cases, the angle α appears to increase as the packing efficiency. Further, the plane projections of these trigons/rhombohedrons clearly provide two distinct pictures (Figs. 6 and 7), first case shows doubly centered while the other two show simple (primitive) hexagonal unit cell. The results led us to conclude the distinctness of Trigon and RCP.

In the text, we have described in detail the formation of HCP and CCP, taking pace along [001] direction. This direction is clearly different from the direction [111], the direction of close packing in FCC. The two forms, i.e. CCP and FCC are indistinguishable in elements while in MX system (for example, ZnS, SiC and NaCl), they are clearly distinguishable. Further, all polytypic transformations are explained on the basis of layer displacement mechanism, which are taking place in the same horizontal plane even in between HCP and CCP and no change of direction is taking place during or after the transformation is complete. Thus, the distinctness of CCP and FCC should be taken into account and consequently the existing ambiguities and confusions related to them should be treated as removed.

Now, we know that RCP unit is the primitive rhombohedron constructed inside a CCP unit and represents one extreme form of close packing. Similarly, HCP represents another extreme form of close packing. If they are not considered independent space lattices, the close packing will have no representation at all because associating HCP with SH and RCP (CCP) with SC, both (SC and SH) non close packed structures is extremely unjustified. They need to be treated separately as the two forms of extreme close packing.

As far as the proper assignment of crystal systems to HCP and RCP lattices is concerned, this is to suggest that based on their similar unit cell shapes and symmetries, they should continue with hexagonal crystal system and trigonal crystal system respectively but as exclusive and independent members. Further, as we have seen above that the rhombohedral unit is contained within the CCP/FCC unit, it also inherits and possesses all symmetries of CCP/FCC as illustrated in all 5 cubic point groups (viz. 23,432, m3, $\bar{4}3m$ and m3m) shown in Fig. 8. However, carefully examining the point group symmetries of HCP unit (Fig. 9), we find it is $\bar{6}m2$ and not 6mm or 6/mmm. On the other hand, on the basis of the latter two point symmetries, the space groups of HCP unit have been reported in the literature(9,10). Accordingly, the necessary corrections are needed for space group representation.

The consideration of HCP and RCP as exclusive and independent lattices helps to remove the discrepancy existing in the relationships between:

1) the lattice types and point groups in 3 dimensions (Table 1)
2) the lattice types and symmorphic space groups (Table 2)

Based on the above changes, the original Table and Figures representing 14 Bravais (space) lattices will get modified. They are respectively provided in Table 5 and Fig.10.

## 8. Conclusion

The complicacies, confusions and ambiguities encountered in the representation of trigonal/rhombohedral and hexagonal lattices on one hand and the FCC and CCP lattices on the other hand have been removed by suggesting the need to consider two more (HCP and RCP) independent lattices with proper justifications. Their inclusion very well explains the correct relationships that should exist between the number of lattices with the point groups and the symmorphic space groups, respectively. The above finding suggests the need to correct the wrongly assigned symmetry (and hence the corresponding space groups) to HCP and CCP so that the observed hexagonal – cubic transformations (specially in close packed polytypic compounds) could be explained crystallographically as well. This finding will certainly have immediate as well as far reaching implications.

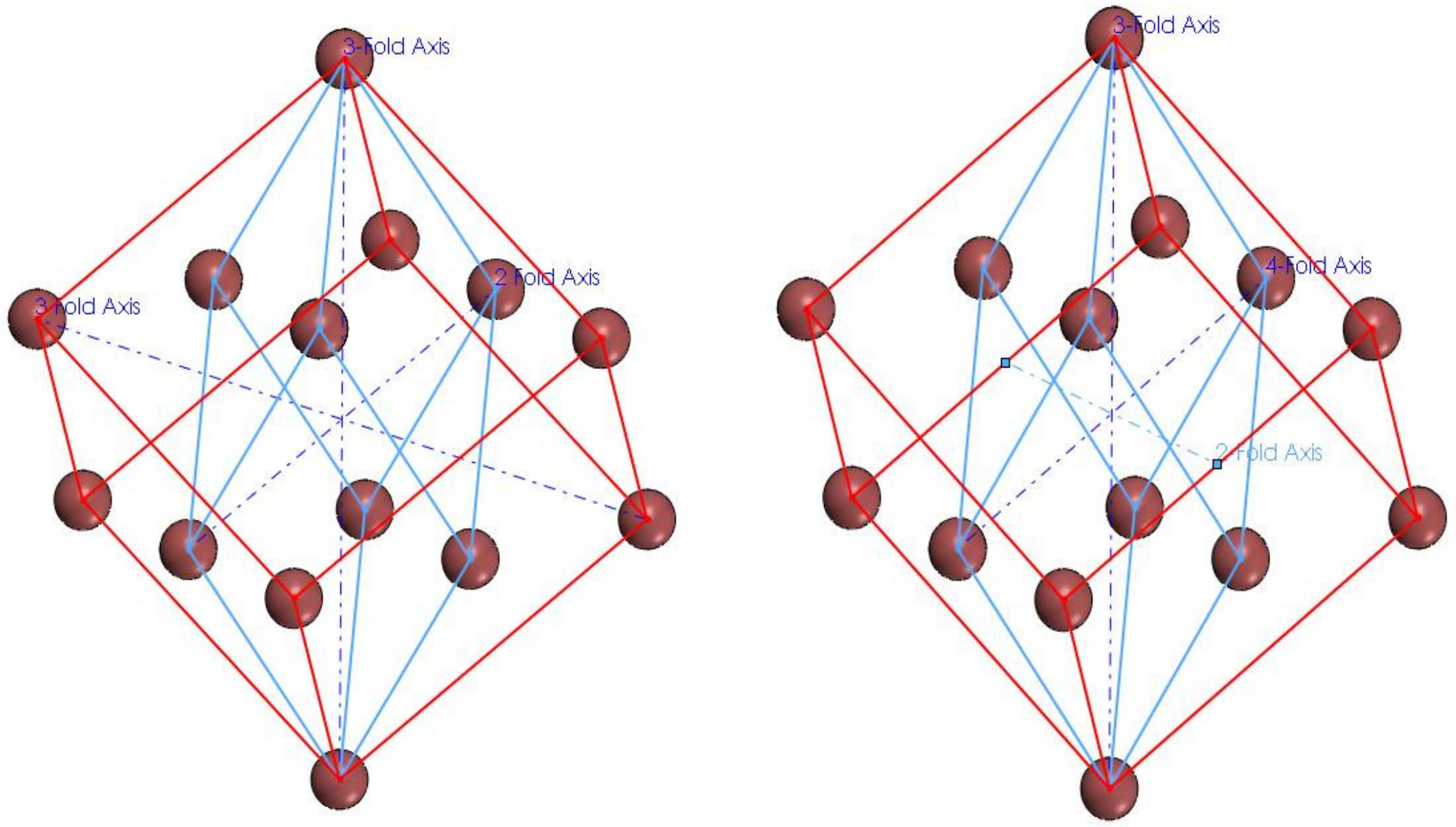

(a) (b)

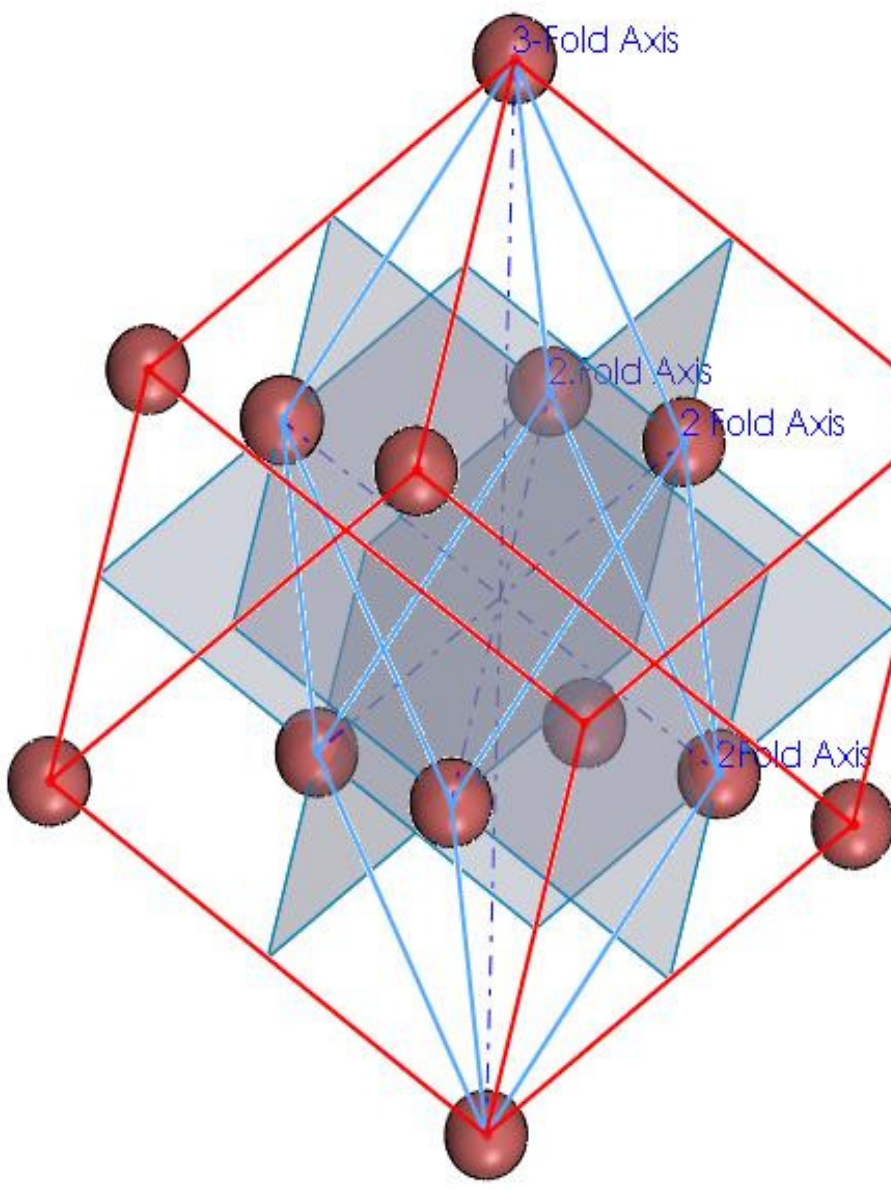

(c)

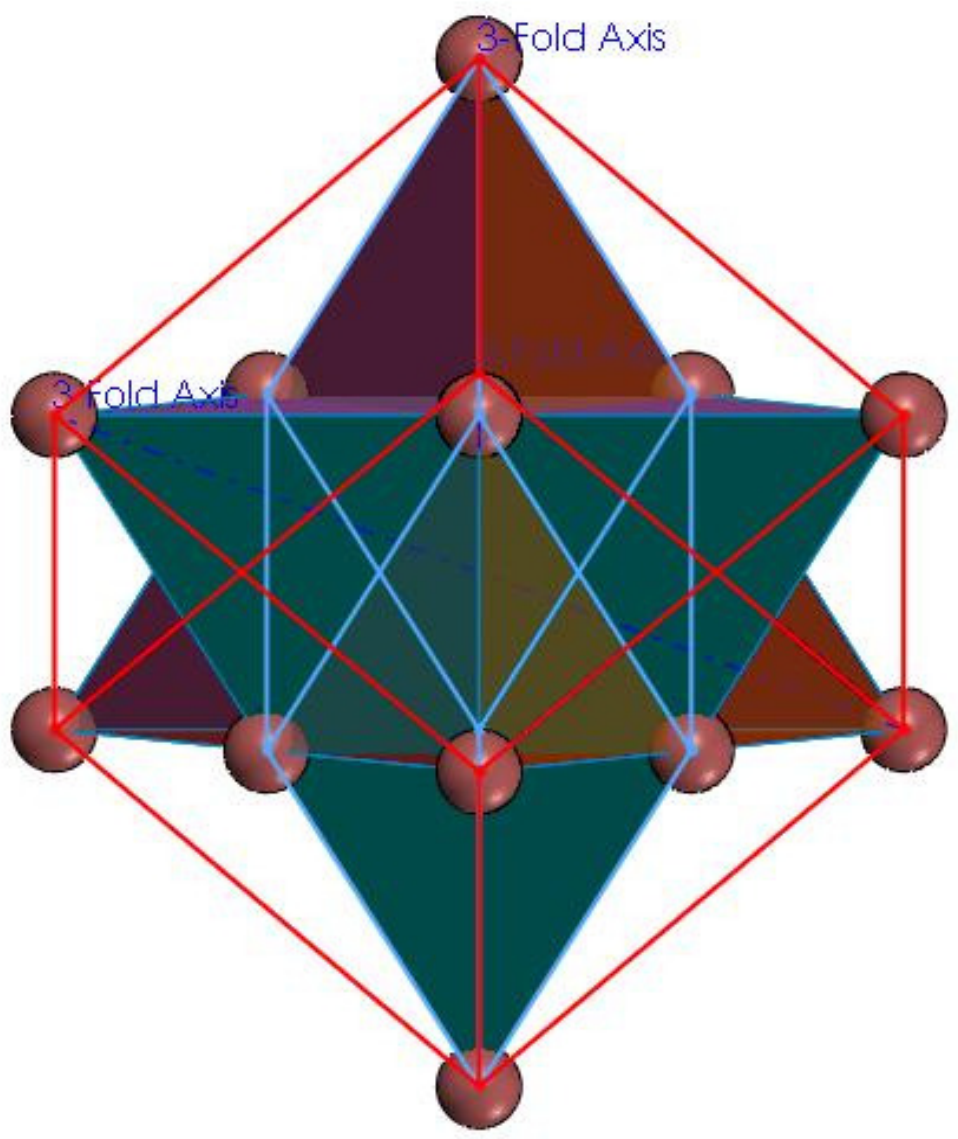

(d)

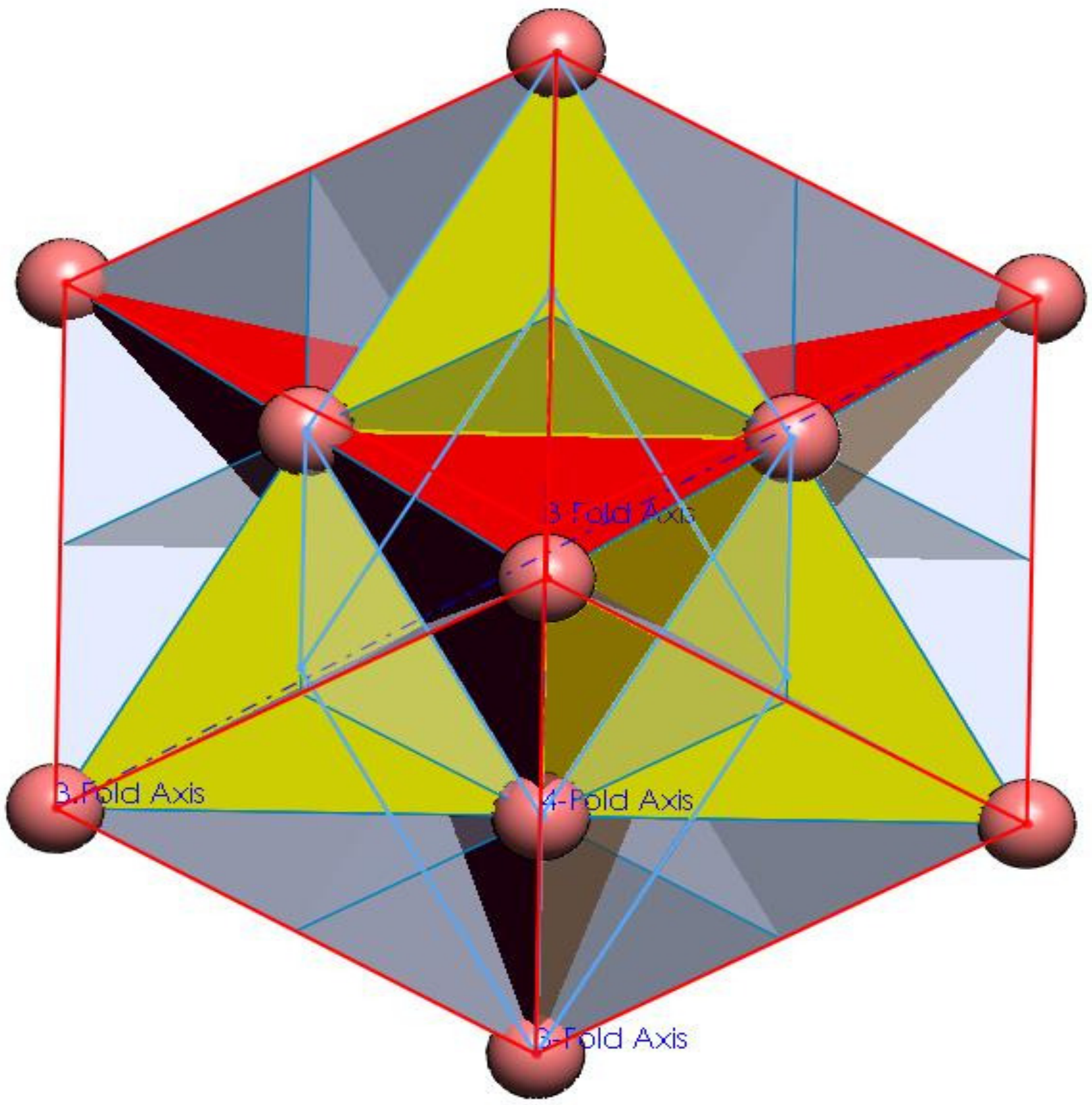

(e)

Fig. 8 Showing 5 point group symmetries of a cube, also possessed by an RCP unit cell

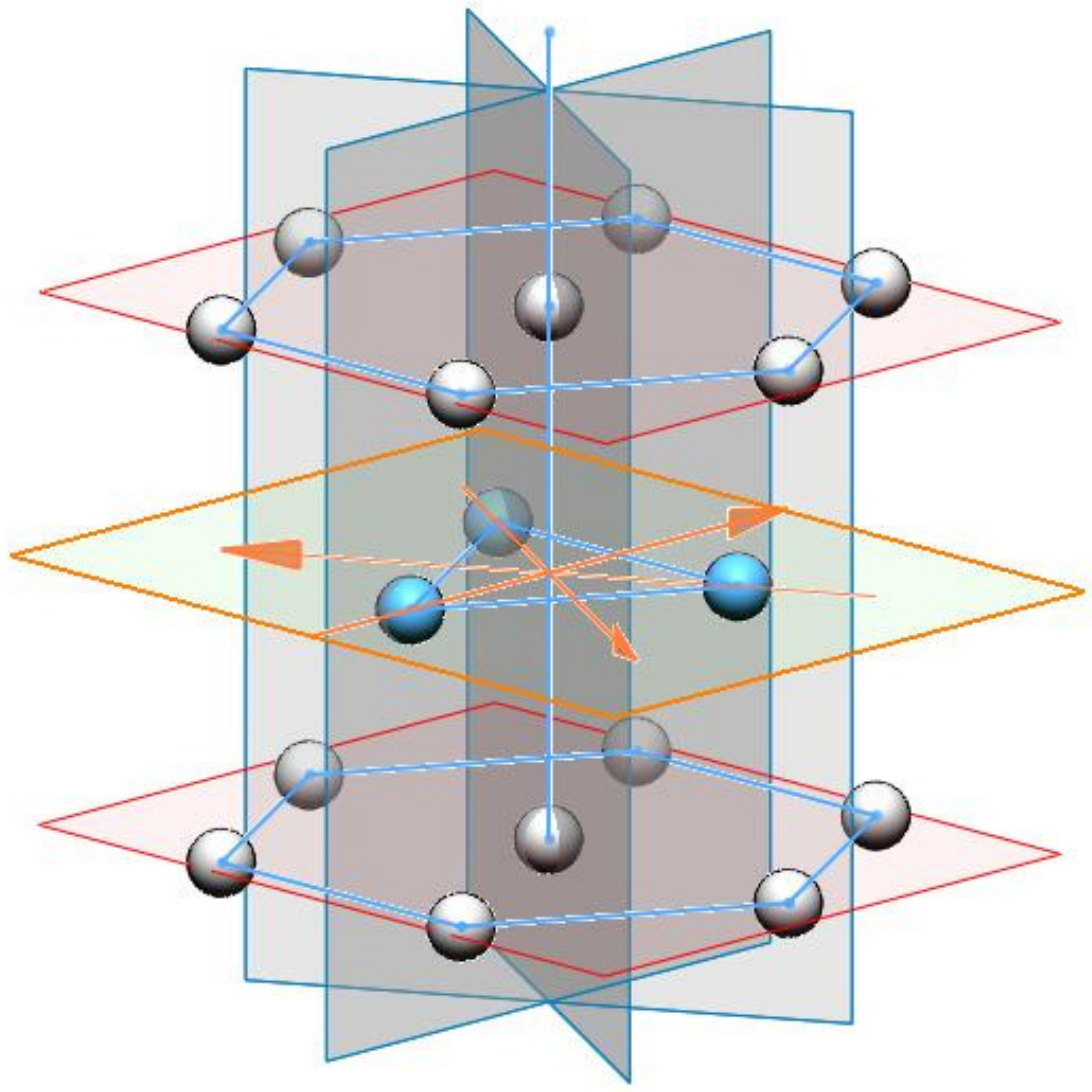

Fig. 9 Showing the point group symmetries possessed by an HCP unit cell

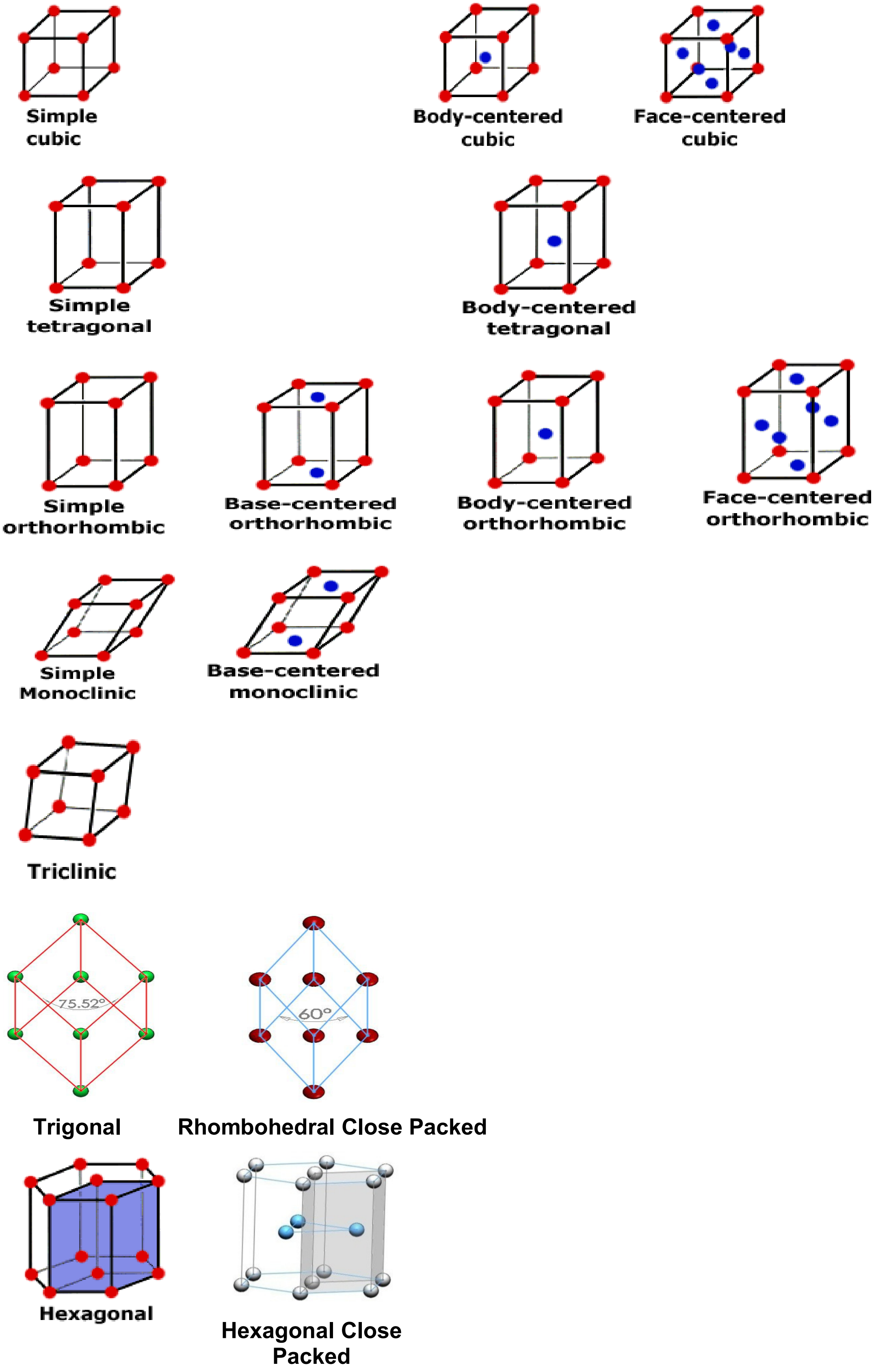

Fig. 10 Showing 16 BW (Bravais-Wahab) lattices

Table 1. Crystallographic data for one, two and three dimensions

| Dimension | Lattice Types | | Point Groups | Symmorphic Space Groups | |
|---|---|---|---|---|---|
| | Previous | Revised | | Previous | Revised |
| One | 1 | 1 | 2 | 2 | 2 |
| Two | 5 | 5 | 10 | 13 | 13 |
| Three | 14 | 16 | 32 | 61 | 73 |

Table 2. Distribution of symmorphic space groups

| Crystal System | No. of Point Groups | Lattice Types | | Symmorphic Space Groups | |
|---|---|---|---|---|---|
| | | Previous | Revised | Previous | Revised |
| Cubic | 5 | 3 | 3 | 15 | 15 |
| Tetragonal | 7 | 2 | 2 | 14 | 14 |
| Orthorhombic | 3 | 4 | 4 | 12 | 12 |
| Monoclinic | 3 | 2 | 2 | 6 | 6 |
| Triclinic | 2 | 1 | 1 | 2 | 2 |
| Hexagonal | 7 | 1 | 2 | 7 | 14 |
| Trigonal | 5 | 1 | 2 | 5 | 10 |
| Total | 32 | 14 | 16 | 61 | 73 |

Table 3. Angle and efficiency of trigon/rhombohedron in some crystal systems

| Crystal System | Calculated Angle of Trigon/Rhombohedron | Efficiency (%) |
|---|---|---|
| CCP (FCC) | 60.0 | 74 |
| SH | 75.5 | 60 |
| BCC | 109.47 | 68 |

Table 4. List of low period polytypes

| Polytype | Hexagonal ABC-sequence | Polytype | Rhombohedral ABC-sequence | Remark |
|---|---|---|---|---|
| 1H | A/A | 3R (∞) | A│B│C/A | Obverse |
| | | | A│C│B/A | Reverse |
| 2H(11) | AB/A | 6R (∞) | AB│CA│BC/A | Obverse |
| | AC/A | | AC│BA│CB/A | Reverse |
| 3H | − | 9R $(21)_3$ | ABC│BCA│CAB/A | Obverse |
| | | $(12)_3$ | ABA│CAC│BCB/A | Reverse |
| 4H (22) | ABCB/A | 12R $(31)_3$ | ABCA│CABC│BCAB/A | Obverse |
| | ABAC/A | $(13)_3$ | ABAC│BCBA│CACB/A | Reverse |
| 5H (41) | ABCAB/A | 15R $(32)_3$ | ABCAC│BCABA│CABCB/A | Obverse |
| | | $(23)_3$ | ABCBA│CABAC│BCACB/A | Reverse |

Table 5. Space Lattices in 3 Dimensions

| Crystal System | Cell Axes and Angles | Associated Lattice | | Characteristic Symmetry Elements | To be Specified | | |
|---|---|---|---|---|---|---|---|
| | | Number | Symbol* | | Axes | Angles | Total Parameters |
| Triclinic | a≠b≠c α≠β≠γ≠90° | 1 | P | None | a,b,c | α,β,γ | 6 |
| Monoclinic | a≠b≠c α=γ=90°≠β | 2 | P,C | $1(2/\bar{2} \equiv m)$ | a,b,c | γ | 4 |
| Ortho rhombic | a≠b≠c α=β=γ=90° | 4 | P,C,I,F | $3(2/\bar{2} \equiv m)$ | a,b,c | ----- | 3 |
| Tetragonal | a=b≠c α=β=γ=90° | 2 | P,I | $1(4/\bar{4})$ | a,c | ----- | 2 |
| Trigonal / RCP | a=b=c (for both) 60°<α<120° α≠90° / α=60° | 2 | P,RCP | $1(3/\bar{3})$ | a | α | 2 |
| Hexagonal HCP | a=b≠c α=β=90° γ=120° | 2 | P,HCP | $1(6/\bar{6})$ | a,c | ----- | 2 |
| Cubic | a=b=c α=β=γ=90° | 3 | P,I,F | $4(3/\bar{3})$ | a | ----- | 1 |

* **Lattice Types**

P – Primitive (Lattice points are at the corners of the unit cell only)

C – Side Centered or Base Centered (Lattice points are at the corners and at 2 face centers opposite to each other)

I – Body Centered (Lattice points are at the corners and at the body center)

F – Face Centered (Lattice points are at the corners and at the 6 face centers)

HCP – Hexagonal Close Packing (Lattice points are at the corners and at $\frac{2}{3}\frac{1}{3}\frac{1}{2}$)

RCP--Rhombohedral Close Packing (Lattice points are at the corners and at $\frac{2}{3}\frac{1}{3}\frac{1}{2}$ and $\frac{1}{3}\frac{2}{3}\frac{1}{2}$)